# Floquet nonadiabatic dynamics in open quantum systems

Vahid Mosallanejad[1,2*], Yu Wang[1,2], Jingqi Chen[1,2], Wenjie Dou[1,2,3*]

[1] *Department of Chemistry, School of Science, Westlake University, Hangzhou, Zhejiang 310024, China*

[2] *Institute of Natural Sciences, Westlake Institute for Advanced Study, Hangzhou, Zhejiang 310024, China*

[3] *Department of Physics, School of Science, Westlake University, Hangzhou 310024 Zhejiang, China*

Email: vahid@westlake.edu.cn; douwenjie@westlake.edu.cn

The Born–Oppenheimer (BO) approximation has shaped our understanding on molecular dynamics microscopically in many physical and chemical systems. However, there are many cases that we must go beyond the BO approximation, particularly when strong light-matter interactions are considered. Floquet theory offers a powerful tool to treat time-periodic quantum systems. In this overview, we briefly review recent developments on Floquet nonadiabatic dynamics, with a special focus on open quantum systems. We first present the general Floquet Liouville von-Neumann (LvN) equation. We then show how to connect Floquet operators to real time observables. We proceed to outline the derivation of the Floquet quantum master equation in treating the dynamics under periodic driving in open quantum systems. We further present the Floquet mixed quantum classical Liouville equation (QCLE) to deal with coupled electron-nuclear dynamics. Finally, we embed FQCLE into a classical master equation (CME) to deal with Floquet nonadiabatic dynamics in open quantum systems. The formulations are general platforms for developing trajectory based dynamical approaches. As an example, we show how Floquet QCLE and Floquet CME can be implemented into a Langevin dynamics with Lorentz force and surface hopping algorithms.

## 1. INTRODUCTION

Molecular dynamics (MD) in the presence of an external periodic driving field is a research topic of intense interests both experimentally and theoretically [1, 2, 3, 4]. For instance, using laser as the driving field, phenomena such as higher-order harmonic generation, molecular structural deformations, photodissociation and ionization have been observed [5, 6]. Early MD simulations were based on the Born-Oppenheimer (BO) or adiabatic approximation which simply implies that nuclei classically evolve on one Potential Energy Surface (PES), typically the ground state PES [7]. However, growing evidence supports the breakdown of the BO approximation in many systems [8, 9]. For instance, the appearance of non-adiabatic coupling is evident in a molecular junction subjected to an intense laser light [10]. In general, non-adiabatic phenomena can happen during: (I) excited state dynamics, and (II) in presence of external driving [11]. Varieties of theoretical approaches such as time-dependent Schrödinger equation (which is practical only for small systems like $H_2^+$), multi-configurational time dependent Hartree (MCTDH) [12], hierarchical quantum master equation (HQME) [13,14], exact factorization [15], Ab Initio Multiple Spawning [16], decoherence corrected fewest-switches surface hopping [17, 18], and other methods are proposed to incorporate nonadiabatic effects into the molecular dynamics. Currently, developing strategies that can incorporate a periodic external driving to the current non-adiabatic MD methods are one of the main research tasks [19, 20, 21].

The modest aim of this Overview is to highlight how the mixed Quantum-Classical Liouville equation (QCLE), and QCLE-classical master equation (QCLE-CME) can be complemented by the Floquet theorem to deal with nonadiabatic dynamics with periodic driving in closed and open quantum systems. Floquet theory was first introduced by Gaston Floquet in 1883, when he analysed the stability of a first order linear partial differential equation (PDE), $\dot{X}(t) = A(t)X(t)$, provided that the matrix $A(t)$ is periodic [22]. Floquet theorem thus provides a method for the analysis of dynamical systems subjected to a periodic driving. Since Schrödinger equation is a first order time derivative PDE, the Floquet theorem is applicable for variety of quantum systems with time-periodic Hamiltonians. Nearly 80 years later, Shirley introduced Floquet time-evolution operator to solve the time-periodic Schrödinger equation [23]. The Floquet theorem for a time-periodic Hamiltonian is analogous to the Bloch's theorem for position-periodic Hamiltonians in solids [24,25]. Floquet engineering in the solid-state usually results in splitting of the Bloch bands (a hybrid state often called Floquet-Bloch's states or Floquet quasi-energy spectrum) [26, 27], and in many cases such driven systems support nontrivial topology [28]. Knowing BO approximation breaks when electronic states getting close in energy (nontrivial avoid crossing) and the PES in molecular dynamics is the analogues concept to the Bloch's bands in solid, one may get this impression that equivalent Floquet PES (replicas) can gives us a powerful tool to deal with non-adiabatic dynamics in presence of intense light.

This paper is structured as follows. In section 2, we will first introduce the Floquet Liouville von-Neumann equation. Then, we explain two types of Floquet Redfield quantum master equations (QME) to deal with Floquet dynamics in open quantum systems. In section 3, we will introduce the Floquet quantum classical Liouville equation (QCLE) to incorporate nuclear motion classically. Then, we will introduce the Floquet electronic friction and its numerical realization using the concept of one-body Floquet Green's function. In section 4, we introduce our Floquet QCLE-CME approaches to deal with nonadiabatic dynamics in open quantum systems in general. Then, we employ our surface hopping algorithms to solve this Floquet QCLE-CME for a time-periodic Anderson-Holstein model.

## 2. FLOQUET QUANTUM MASTER EQUATION

### 2.1 Fourier and Floquet representations of Liouville von-Neumann (LvN) equation

In the Schrödinger picture, the Liouville von-Neumann (LvN) equation is

$$\frac{d}{dt}\hat{\rho}(t) = -i[\hat{H}(t), \hat{\rho}(t)]. \tag{1}$$

We have set $\hbar = 1$. $\hat{\rho}(t)$ is the total density operator that includes the system and the bath. $\hat{H}(t)$ is the total Hamiltonian that is time-periodic, $\hat{H}(t + T) = \hat{H}(t)$. A convenient and elegant way to derive time evolution of a time-periodic quantum system is based on a sequence of three closely related representations of LvN equation as: (I) The Fourier vector-like (V-like) representation, (II) Fourier matrix-like (M-like) representation and (III) Floquet M-like representation. In the first representation, we substitute the following Fourier expansions

$$\hat{\rho}(t) = \sum_n \hat{\rho}^{(n)}(t) e^{in\omega t}, \tag{2}$$

$$\hat{H}(t) = \sum_n \hat{H}^{(n)} e^{in\omega t}, \tag{3}$$

into LvN equation:

$$\sum_n \frac{d\hat{\rho}^{(n)}(t)}{dt} e^{in\omega t} + in\omega \hat{\rho}^{(n)}(t) e^{in\omega t} = \sum_{n,m} -i[\hat{H}^{(n-m)}, \hat{\rho}^{(m)}(t)] e^{in\omega t}. \tag{4}$$

We call the above expression as the Fourier V-like representation of LvN equation.

To obtain the type (II), Fourier M-like representation, we now introduce ladder operator $\hat{L}_n$, which is an operator in the Flourier space $\{|n\rangle\}$ with following properties

$$\hat{L}_n |m\rangle = |m + n\rangle, \qquad \hat{L}_n \hat{L}_m = \hat{L}_m \hat{L}_n = \hat{L}_{n+m} \Longrightarrow [\hat{L}_n, \hat{L}_m] = 0. \tag{5}$$

Such that we can define the density operator and Hamiltonian in the Fourier M-like representation

$$\hat{\rho}^f(t) = \sum_n \hat{L}_n \otimes \hat{\rho}^{(n)}(t) e^{in\omega t}, \tag{6}$$

$$\hat{H}^f(t) = \sum_n \hat{L}_n \otimes \hat{H}^{(n)} e^{in\omega t}. \tag{7}$$

We expect that the LvN equation also holds in the Fourier M-like representation

$$\frac{d}{dt}\hat{\rho}^f(t) = -i[\hat{H}^f(t), \hat{\rho}^f(t)]. \tag{8}$$

Because if we substitute $\hat{\rho}^f(t)$ and $\hat{H}^f(t)$ in the LvN, we arrive at

$$\sum_n \hat{L}_n \otimes \frac{d\hat{\rho}^{(n)}(t)}{dt} e^{in\omega t} + in\omega \hat{L}_n \otimes \hat{\rho}^{(n)}(t) e^{in\omega t} = \sum_{n,m} -i\hat{L}_n \otimes [\hat{H}^{(n-m)}, \hat{\rho}^{(m)}(t)] e^{in\omega t}, \tag{9}$$

which is the corresponding Fourier M-like representation of LvN in Fourier V-like representation.

To arrive at the Floquet M-like presentation, we define

$$\hat{\rho}^F(t) = e^{-i\hat{N}\otimes\mathbf{1}\omega t}\ \hat{\rho}^f(t)\ e^{i\hat{N}\otimes\mathbf{1}\omega t}. \tag{10}$$

Here, $\hat{N}$ is the number operator in the Fourier space with the following properties:

$$\hat{N}|n\rangle = n|n\rangle, \qquad [\hat{N}, \hat{L}_n] = +n\hat{L}_n. \tag{11}$$

Such that we can rewrite the density operator in the Floquet M-like representation as

$$\hat{\rho}^F(t) = \sum_n e^{-i\hat{N}\otimes\mathbf{1}\omega t}\hat{L}_n \otimes \hat{\rho}^{(n)}(t) e^{i\hat{N}\otimes\mathbf{1}\omega t} e^{in\omega t} = \sum_n \hat{L}_n \otimes \hat{\rho}^{(n)}(t). \tag{12}$$

In the above equation, we have used Baker-Campbell-Hausdorff (BCH). Notice that the phase factor $e^{in\omega t}$ is vanished. We find that $\hat{\rho}^F(t)$ follows the LvN equation in the Floquet representation:

$$\frac{d}{dt}\hat{\rho}^F(t) = -i[\hat{N}\otimes\mathbf{1}\omega, \hat{\rho}^F(t)] - i\left[\sum_n \hat{L}_n \otimes H^{(n)}, \hat{\rho}^F(t)\right] \equiv -i[\hat{H}^F, \hat{\rho}^F(t)]. \tag{13}$$

Here we define the Floquet Hamiltonian as:

$$\hat{H}^F = \sum_n \hat{L}_n \otimes H^{(n)} + \hat{N}\otimes\mathbf{1}\omega \tag{14}$$

The above equation is what we refer to as the Floquet LvN equation. We have reduced the time-dependent Hamiltonian into a time-independent Hamiltonian. As a result, the Floquet basis is a tensor product of the Hilbert-space basis $\{|\alpha\rangle\}$ and the Fourier basis $\{|n\rangle\}$: $\{|\alpha, n\rangle\} = \{|n\rangle\} \otimes \{|\alpha\rangle\}$. The form of $\hat{H}^F$ is consistent with the form that is obtained from the expansion of the time-evolution operator or the form that is derived from the concept of quasi-energy [23, 24]. However, in our derivation, we arrive at the Floquet LvN equation that is formally equivalent to the non-Floquet LvN equation, which will allow us to derive formal results that will show below.

**2.2 Evolution in Floquet representation**

Since the Floquet Hamiltonian is time-independent, we can formally solve the Floquet LvN equation:

$$\hat{\rho}^F(t) = \hat{U}^F(t, t_0)\hat{\rho}^F(t_0)\hat{U}^{F^{-1}}(t, t_0), \qquad \hat{U}^F(t, t_0) = e^{-i\hat{H}^F(t-t_0)}. \tag{15}$$

$\widehat{U}^F(t, t_0)$ is the Floquet time evolution operator. With this solution, we can calculate observables. For any operator, we can define the corresponding operator in Floquet M-like representation,

$$\widehat{O}^F = \sum_k \widehat{L}_k \otimes \widehat{O}^{(k)} \tag{16}$$

Here $\widehat{O}^{(k)}$ is the corresponding Fourier component, $\widehat{O}(t) = \sum_k \widehat{O}^{(k)} e^{ik\omega t}$. The expectation value of observable, $\langle \widehat{O}(t) \rangle$ can be expressed as

$$\langle \widehat{O}(t) \rangle = \sum_{k,n} Tr(\widehat{O}^{(k)} \widehat{\rho}^{(n)}(t)) e^{i(n+k)\omega t} = \sum_{m,n} Tr(\widehat{O}^{(m-n)} \widehat{\rho}^{(n)}(t)) e^{im\omega t}. \tag{17}$$

Here, $Tr$ denotes trace over Hilbert space. Notice that in the Floquet M-like representation, we have

$$\widehat{O}^F \widehat{\rho}^F(t) = \sum_{k,n} \widehat{L}_k \otimes \widehat{O}^{(k)} \widehat{L}_n \otimes \widehat{\rho}^{(n)}(t) = \sum_{k,n} \widehat{L}_k \widehat{L}_n \otimes \widehat{O}^{(k)} \widehat{\rho}^{(n)}(t) = \sum_m \widehat{L}_m \otimes \sum_n \widehat{O}^{(m-n)} \widehat{\rho}^{(n)}(t), \tag{18}$$

Such that the observable can be calculated as

$$\langle \widehat{O}(t) \rangle = \sum_m Tr(\langle m | \widehat{O}^F \widehat{\rho}^F(t) | 0 \rangle) e^{im\omega t} \tag{19}$$

Here, we have used the property of the Ladder operator, $\langle m | \widehat{L}_m | 0 \rangle = 1$.

Similarly, the time evolution operator in Schrödinger representation $\widehat{U}(t, t_0)$ can be expressed in terms of its Floquet version, $\widehat{U}^F(t, t_0)$

$$\widehat{U}(t, t_0) = \sum_k \widehat{U}^{(k)}(t, t_0) e^{ik\omega t} = \sum_k \langle k | \widehat{U}^F(t, t_0) | 0 \rangle e^{ik\omega t} = \sum_k \langle k | e^{-i\widehat{H}^F(t-t_0)} | 0 \rangle e^{ik\omega t}. \tag{20}$$

The above matrix form for $\widehat{U}(t, t_0)$ is identical to the relation achieved from the expansion solution [24]. We can also go to the diagonalized (adiabatic) basis set of the $\widehat{H}^F$

$$\widehat{U}(t, t_0) = \sum_k \langle k | YY^\dagger e^{-i\widehat{H}^F(t-t_0)} YY^\dagger | 0 \rangle e^{ik\omega t} = \sum_k \langle k | Y e^{-i\widehat{\Lambda}^F(t-t_0)} Y^\dagger | 0 \rangle e^{ik\omega t}. \tag{21}$$

$Y$ and $\widehat{\Lambda}^F$ are the eigenvectors and eigenvalues of the Floquet Hamiltonian such that $Y^\dagger \widehat{H}^F Y = \widehat{\Lambda}^F$.

**2.3 Redfield QME**

Quantum master equation (QME) is the one of the simplest way to treat dynamics of open systems [29]. Since the focus of this overview is on Floquet dynamics in open quantum systems, we will first cover the Floquet QME. A typical open quantum system consists of a small size system (the System) and multiple rather large-size terminals (the bath) with very dense non-interaction states in thermodynamic equilibrium. Here, we will identify the total Hamiltonians of as System $\hat{H}_S$ and Bath $\hat{H}_B$. In addition, $\hat{H}_{BS}$ denotes the interaction of the System with environment.

In the followings we intend to explain two forms for Floquet QME. In general, QME enables us to study the time-evolution of the System's density operator $\rho_S(t)$ (so-called the reduced density operator) rather than the untraceable total density operator $\rho_{tot}(t)$. From the $\rho_S(t)$, one can evaluate other physical observables such as the expectation value of Number operator. In what follows, we assume that only the System's Hamiltonian is time-periodic $H_s(t) = H_s(t + T)$, in which $T = 2\pi/\omega$, and $\omega$ are the driving period and the driving frequency. With this restriction, it appears to us that there could be two approaches in which the QME can benefit from the mathematical Floquet theory. In fact, this first approach is the Redfield EOM for the reduced $\rho_S(t)$ wherein the Floquet theorem simplifies the solution procedure of the EOM. In the second approach, we will first transform the total density operator to the Floquet representation as $\hat{\rho}_{tot}(t) \to \hat{\rho}_{tot}^F(t)$ and then the result will be reduced into the System's Floquet density matrix as $\rho_{tot}^F(t) \to \rho_S^F(t)$.

Using the second quantization form, a general expression for the three major components of the total Hamiltonian can be given by:

$$\hat{H}_S(t) = \sum_{\alpha,\beta} h_{\alpha\beta}(t)\, \hat{c}_\alpha^\dagger \hat{c}_\beta, \qquad \hat{H}_B = \sum_{k \in l,r} \varepsilon_k\, \hat{c}_k^\dagger \hat{c}_k, \qquad \hat{H}_{BS} = \sum_{k \in l,r; \alpha} V_{k\alpha}^* \hat{c}_\alpha^\dagger \hat{c}_k + V_{k\alpha} \hat{c}_k^\dagger \hat{c}_\alpha, \qquad (22)$$

where $\hat{c}_\alpha^\dagger$ is the creation operator for an electronic orbital in the system and $\hat{c}_k^\dagger$ is the creation operator for electronic orbital $k$ in the bath. $h_{\alpha\beta}$ is a matrix element of the system Hamiltonian. We have set that the system is time-periodic $H_s(t) = H_s(t + T)$. For the system-bath couplings, we employ the wide-band approximation [30] such that the hybridization function $\Gamma_{\alpha\beta}(\varepsilon)$ is independent of $\varepsilon$

$$\Gamma_{\alpha\beta}(\varepsilon) = 2\pi \sum_k V_{k\alpha} V_{k\beta}^* \delta(\varepsilon - \varepsilon_k) = \Gamma_{\alpha\beta}. \qquad (23)$$

To derive the Redfield QME with periodic driving, we start with the LvN equation for the total density operator in the interaction picture,

$$\frac{d}{dt}\tilde{\rho}_{tot}(t) = -i\big[\widetilde{H}_{BS}(t), \tilde{\rho}_{tot}(t)\big]. \qquad (24)$$

Here the total density operator and interaction Hamiltonian are written in the interaction picture:

$$\widetilde{H}_{BS}(t) = e^{i\hat{H}_B t} \mathcal{T} e^{i\int_0^t \hat{H}_S(s)ds}\, \hat{H}_{BS}\, e^{-i\hat{H}_B t} \mathcal{T} e^{-i\int_0^t \hat{H}_S(s)ds},$$

$$\tilde{\rho}_{tot}(t) = e^{i\hat{H}_B t} \mathcal{T} e^{i\int_0^t \hat{H}_S(s)ds}\, \hat{\rho}_{tot}(t)\, e^{-i\hat{H}_B t} \mathcal{T} e^{-i\int_0^t \hat{H}_S(s)ds}. \qquad (25)$$

where $\mathcal{T}$ is the time-ordering operator. As typically done in the QME, we take the integral over time of the LvN equation and then substitute the result into the LvN equation again, such that we arrive at

$$\frac{d}{dt}\tilde{\rho}_{tot}(t) = -i[\widetilde{H}_{BS}(t),\tilde{\rho}_{tot}(0)] - \int_0^t [\widetilde{H}_{BS}(t),[\widetilde{H}_{BS}(t'),\tilde{\rho}_{tot}(t')]]dt'. \qquad (26)$$

The derivation proceeds by taking the *Born* approximation which states that system and bath are separable due to weak coupling such that the total density matrix can be written as $\tilde{\rho}_{tot}(t) = \tilde{\rho}_S(t) \otimes \tilde{\rho}_B$. We then trace over bath on both sides of the above equation

$$\frac{d}{dt}\tilde{\rho}_S(t) = -\int_0^t Tr_B[\widetilde{H}_{BS}(t),[\widetilde{H}_{BS}(t'),\tilde{\rho}_S(t') \otimes \tilde{\rho}_B]]dt'. \qquad (27)$$

Note that, it can be shown that $Tr_B[\widetilde{H}_{BS}(t),\tilde{\rho}_{tot}(0)] = 0$, provided that $H_{BS}(t)$ is a bi-linear function of the bath and system operators and the initial density operator is in equilibrium. The last assumption is the *Markovian* approximation, in which one assumes that the bath dynamics are much faster than the system, such that one can ignore the memory effects of the bath dynamics:

$$\frac{d}{dt}\tilde{\rho}_S(t) = -\int_0^\infty Tr_B[\widetilde{H}_{BS}(t),[\widetilde{H}_{BS}(t-\tau),\tilde{\rho}_S(t) \otimes \tilde{\rho}_B]]\, d\tau. \qquad (28)$$

The above equation is the Redfield QME that is true with or without time-dependent driving [31]. Below, we will simplify the QME when applying it to our model Hamiltonian with Floquet driving.

### 2.3.1 Redfield QME with Floquet driving: Hilbert space QME

We now go back to the Schrödinger picture, where the density operator in Schrödinger picture reads $\hat{\rho}_S(t) = \widehat{U}_S(t)\tilde{\rho}_S(t)\widehat{U}_S^\dagger(t)$. Here $\widehat{U}_S(t) = \mathcal{T}e^{-i\int_0^t \widehat{H}_S(s)ds}$ is the time evolution operator for the system. Also, we define $\widehat{U}_S(t, t-\tau) = \mathcal{T}e^{-i\int_{t-\tau}^t \widehat{H}_S(s)ds}$ and $\widetilde{\widetilde{H}}_{SB}(t,\tau) = \widehat{U}_B(\tau)\widehat{U}_S(t,t-\tau)\widehat{H}_{BS}\widehat{U}_S^\dagger(t,t-\tau)\widehat{U}_B^\dagger(\tau)$. With these definitions, the Redfield QME in the Schrödinger picture is given by

$$\frac{\partial \hat{\rho}_S(t)}{\partial t} = -i[\widehat{H}_S(t),\hat{\rho}_S(t)] - \int_0^\infty Tr_B[\widehat{H}_{BS},[\widetilde{\widetilde{H}}_{SB}(t,\tau),\hat{\rho}_S(t) \otimes \hat{\rho}_B]]\, d\tau, \qquad (29)$$

The bilinear form of the system-bath couplings allows us to simplify $\widetilde{\widetilde{H}}_{SB}(t,\tau)$ as

$$\widetilde{\widetilde{H}}_{SB}(t,\tau) = \sum_\alpha \widetilde{\widetilde{B}}_\alpha^\dagger(\tau)\tilde{\tilde{c}}_\alpha(t,\tau) + \tilde{\tilde{c}}_\alpha^\dagger(t,\tau)\widetilde{\widetilde{B}}_\alpha(\tau). \qquad (30)$$

Here, $\widetilde{\widetilde{B}}_\alpha(\tau) = \sum_k V_{k\alpha}^* \hat{c}_k\, e^{i\varepsilon_k \tau}$, and $\tilde{\tilde{c}}_\alpha(t,\tau) = \widehat{U}_S(t,t-\tau)\hat{c}_\alpha \widehat{U}_S^\dagger(t,t-\tau)$. Same definition holds for $\tilde{\tilde{c}}_\alpha^\dagger(t,\tau)$. After breaking the commutators in Equation (29), we will have 8 nonvanishing terms. We now focus on the following dissipation operator:

$$\int_0^\infty \sum_{\alpha,\beta,k} V_{k\alpha} V_{k\beta}^* f(\varepsilon_k,\mu) e^{i\varepsilon_k \tau} \hat{c}_\alpha \tilde{\tilde{c}}_\beta^\dagger(t,\tau) \hat{\rho}_S(t) \, d\tau. \tag{31}$$

On above, we have used the fact that $Tr_B\left(\hat{c}_k^\dagger \hat{c}_k \hat{\rho}_B(\mu)\right) = f(\varepsilon_k, \mu)$. At this point, we use the results from Section 2.2 for the time evolution, such that we can simplify $\tilde{\tilde{c}}_\beta^\dagger(t,\tau)$ as

$$\tilde{\tilde{c}}_\beta^\dagger(t,\tau) = \sum_{n,m} \langle n|e^{-i\hat{H}^F \tau} \hat{\mathbb{C}}_{\beta,00}^\dagger e^{i\hat{H}^F \tau}|m\rangle e^{i(n-m)\omega t}. \tag{32}$$

Here, we have defined $\hat{\mathbb{C}}_{\beta,00}^\dagger = |0\rangle \hat{c}_\beta^\dagger \langle 0|$ in the Floquet M-like representation. To proceed, we then use the eigenbasis of Floquet system Hamiltonian $Y^\dagger \hat{H}^F Y = \hat{\Lambda}^F$. In particular, we will employ the matrix element

$$\left(Y^\dagger e^{-i\hat{H}^F \tau} \hat{\mathbb{C}}_{\beta,00}^\dagger e^{i\hat{H}^F \tau} Y\right)_{NM} = \left(e^{-i\hat{\Lambda}^F \tau} Y^\dagger \hat{\mathbb{C}}_{\beta,00}^\dagger Y e^{i\hat{\Lambda}^F \tau}\right)_{NM} = \left(Y^\dagger \hat{\mathbb{C}}_{\beta,00}^\dagger Y\right)_{NM} e^{-i(E_N - E_M)\tau}. \tag{33}$$

With this matrix elements, we can evaluate the integral

$$\sum_k V_{k\alpha} V_{k\beta}^* f(\varepsilon_k,\mu) \int_0^\infty e^{i\varepsilon_k \tau} \left(Y^\dagger e^{-i\hat{H}^F \tau} \hat{\mathbb{C}}_{\beta,00}^\dagger e^{i\hat{H}^F \tau} Y\right)_{NM} d\tau$$
$$= \pi \sum_k V_{k\alpha} V_{k\beta} f(\varepsilon_k,\mu) \left(Y^\dagger \hat{\mathbb{C}}_{\beta,00}^\dagger Y\right)_{NM} \delta(\varepsilon_k - \Omega_{NM}^F) = \frac{\Gamma_{\alpha\beta}}{2} \tilde{\mathbb{C}}_{\beta,00\,NM}^\dagger \tag{34}$$

Here, we have defined $\tilde{\mathbb{C}}_{\beta,00\,NM}^\dagger = \left(Y^\dagger \hat{\mathbb{C}}_{\beta,00}^\dagger Y\right)_{NM} f(\Omega_{NM}^F, \mu)$. If we further define $\tilde{\mathfrak{c}}_\beta^\dagger(t) = \sum_{n,m} \langle n|Y \tilde{\mathbb{C}}_{\beta,00}^\dagger Y^\dagger|m\rangle e^{i(n-m)\omega t}$, we can express the first dissipation operator as $\sum_{\alpha,\beta} (\Gamma_{\alpha\beta}/2) c_\alpha \tilde{\mathfrak{c}}_\beta^\dagger(t) \rho_S(t)$. Finally, the first two dissipators can be given by

$$\frac{\partial \hat{\rho}_S(t)}{\partial t} = -i[\hat{H}_S(t), \hat{\rho}_S(t)]$$
$$- \sum_{\alpha,\beta} \frac{\Gamma_{\alpha\beta}}{2} \left(c_\alpha \tilde{\mathfrak{c}}_\beta^\dagger(t)\rho_S(t) + \hat{c}_\alpha^\dagger \bar{\mathfrak{c}}_\beta(t)\rho_S(t) - c_\alpha \rho_S(t)\tilde{\mathfrak{c}}_\beta^\dagger(t) - \hat{c}_\alpha^\dagger \rho_S(t)\bar{\mathfrak{c}}_\beta(t)\right) + h.c. \tag{35}$$

Where $\bar{\mathfrak{c}}_\beta(t) = \sum_{n,m} \langle n|Y\bar{\mathbb{C}}_{\beta,00} Y^\dagger|m\rangle e^{i(n-m)\omega t}$, with the element definition $\bar{\mathbb{C}}_{\beta,00\,NM} = \left(Y^\dagger \hat{\mathbb{C}}_{\beta,00} Y\right)_{NM} (1 - f(-\Omega_{NM}^F, \mu))$.

### 2.3.2 Redfield QME in Floquet representation: Floquet space QME

Notice that the previous QME was written in Hilbert space, with System Hamiltonian and the dissipators being time dependent. Just as for the total system with periodic driving, we can represent the QME in Floquet basis as well. Introducing, $\hat{\mathbb{C}}_\alpha = \hat{L}_0 \otimes \hat{c}_\alpha$, $\hat{\mathbb{C}}_\alpha^\dagger = \hat{L}_0 \otimes \hat{c}_\alpha^\dagger$, and $\hat{\mathbb{H}}_{BS} = \hat{L}_0 \otimes \hat{H}_{SB}$ the procedure of deriving Floquet Redfield QME in the Schrödinger picture is straightforward

$$\frac{\partial \hat{\rho}_S^F(t)}{\partial t} = -i[\hat{H}_S^F, \hat{\rho}_S^F(t)] - \int_0^\infty Tr_B[\hat{\mathbb{H}}_{SB}, [\tilde{\hat{\mathbb{H}}}_{SB}(\tau), \hat{\rho}_S^F(t) \otimes \hat{\rho}_B]] d\tau \qquad (36)$$

where $\tilde{\hat{\mathbb{H}}}_{SB}(\tau) \equiv e^{-i\hat{H}_B\tau} e^{-i\hat{H}_S^F\tau} \hat{\mathbb{H}}_{SB} e^{i\hat{H}_B\tau} e^{i\hat{H}_S^F\tau}$. We find that Floquet dissipators closely resemble the ones in the time independent case:

$$\frac{\partial \hat{\rho}_S^F(t)}{\partial t} = -i[\hat{H}_S^F, \hat{\rho}_S^F(t)] - \sum_{\alpha\beta} \frac{\Gamma_{\alpha\beta}}{2} (\hat{\mathbb{C}}_\alpha Y \tilde{\mathbb{C}}_\beta^\dagger Y^\dagger \hat{\rho}_S^F(t) + \hat{\mathbb{C}}_\alpha^\dagger Y \bar{\mathbb{C}}_\beta Y^\dagger \hat{\rho}_S^F(t) - \hat{\mathbb{C}}_\alpha \hat{\rho}_S^F(t) Y \tilde{\mathbb{C}}_\beta^\dagger Y^\dagger$$
$$- \hat{\mathbb{C}}_\alpha^\dagger \hat{\rho}_S^F(t) Y \bar{\mathbb{C}}_\beta Y^\dagger) + h.c.$$
(37)

In the above equation, we have defined $\tilde{\mathbb{C}}_{\beta\ NM}^\dagger = (Y^\dagger \hat{\mathbb{C}}_\beta Y)_{NM} f(\Omega_{NM}^F, \mu)$ and $\bar{\mathbb{C}}_{\beta\ NM} = (Y^\dagger \hat{\mathbb{C}}_\beta Y)_{NM}(1 - f(-\Omega_{NM}^F, \mu))$. The above equation is the QME in Floquet representation. FQME offers a simple mean to treat dynamics in open quantum systems with periodic driving. Note that the Floquet space is infinite. In practice, the Floquet replicas are truncated as long as one can converge the results.

## 3. FLOQUET QUANTUM CLASSICAL LIOUVILLE EQUATION (QCLE)

In the above treatment, all DoFs are being treated quantum mechanically. When nuclear DoFs are included in the system, one can derive a mix quantum classical Liouville equation (QCLE) to treat all electronic DoFs quantum mechanically, and all nuclear DoFs classically [32, 33]. The QCLE is the formal starting point in deriving any trajectory based nonadiabatic dynamics methods, including surface hopping and Ehrenfest dynamics [34]. With periodic driving, a Floquet QCLE can be derived as well [11, 35]. The F-QCLE will be the formal equation in developing Floquet surface hopping or Floquet electronic friction methods.

### 3.1 Floquet QCLE

We now consider that there are two subsystems of particles: electrons and nuclei. The total Hamiltonian can be separated into nuclear kinetic energy and electronic Hamiltonian:

$$\hat{H}(\hat{\boldsymbol{R}}, \hat{\boldsymbol{P}}, t) = \sum_\alpha \frac{\hat{P}_\alpha^2}{2M_\alpha} + \hat{V}(\hat{\boldsymbol{R}}, t), \qquad (38)$$

Here $\hat{\boldsymbol{R}} = \{\hat{R}_1, \hat{R}_2, ..., \hat{R}_{3N}\}$ and $\hat{\boldsymbol{P}} = \{\hat{P}_1, \hat{P}_2, ..., \hat{P}_{3N}\}$ are momentum and coordinate operators of the nuclei. $\hat{V}(\hat{\boldsymbol{R}}, t)$ is the electronic Hamiltonian, which is assumed periodic: $\hat{V}(t+T) = \hat{V}(t)$. Since the total Hamiltonian is periodic in time, we can start from the Floquet LvN equation for the total density operator:

$$\frac{\partial \hat{\rho}^F}{\partial t}(\boldsymbol{R}, \boldsymbol{P}, t) = -i[\hat{H}^F(\hat{\boldsymbol{R}}, \hat{\boldsymbol{P}}), \hat{\rho}^F(\hat{\boldsymbol{R}}, \hat{\boldsymbol{P}}, t)] \qquad (39)$$

Similar to the case of non-Floquet QCLE, we proceed to perform the partial Wigner transformation (PWT) with respect to the nuclear coordination on both sides of the above LvN equation. For an operator $\hat{O}$, PWT is given by

$$\hat{O}_W(\boldsymbol{R},\boldsymbol{P},t) \equiv \left(\hat{O}(\hat{\boldsymbol{R}},\hat{\boldsymbol{P}},t)\right)_W \equiv (2\pi)^{-3N} \int_0^\infty d\boldsymbol{Y} e^{-i\boldsymbol{P}\cdot\boldsymbol{Y}} \left\langle \boldsymbol{R}-\frac{\boldsymbol{Y}}{2}\left|\hat{O}(\hat{\boldsymbol{R}},\hat{P},t)\right|\boldsymbol{R}+\frac{\boldsymbol{Y}}{2}\right\rangle. \tag{40}$$

Here, $\boldsymbol{Y}$ is an auxiliary nuclei vector coordinate. The direct consequence of PWT is that the output operator is now a function of nuclei's coordinate vector and momentum vector $(\boldsymbol{R},\boldsymbol{P})$ rather than the operators $(\hat{\boldsymbol{R}},\hat{\boldsymbol{P}})$. Notice that when performing PWT for a product, we have the following relationship:

$$\left(\hat{A}\hat{B}\right)_W(\boldsymbol{R},\boldsymbol{P},t) = \left(\hat{A}\right)_W(\boldsymbol{R},\boldsymbol{P},t) \, e^{-i\overleftrightarrow{\Lambda}/2} \left(\hat{B}\right)_W(\boldsymbol{R},\boldsymbol{P},t). \tag{40}$$

Here, $\Lambda$ denotes the Poisson bracket:

$$\overleftrightarrow{\Lambda} = \sum_\alpha \frac{\overleftarrow{\partial}}{\partial P_\alpha} \frac{\overrightarrow{\partial}}{\partial R_\alpha} - \frac{\overleftarrow{\partial}}{\partial R_\alpha} \frac{\overrightarrow{\partial}}{\partial P_\alpha}. \tag{41}$$

$e^{-i\overleftrightarrow{\Lambda}/2}$ is called the Wigner-Moyal operator. If we keep only the first order of its Tayler expansion $e^{-i\overleftrightarrow{\Lambda}/2} \approx 1 - i\overleftrightarrow{\Lambda}/2$, we will then arrive at the Floquet QCLE:

$$\frac{\partial \hat{\rho}^F{}_W(t)}{\partial t} = -i\left[\hat{V}^F{}_W(\boldsymbol{R},\hat{r},\hat{p}),\hat{\rho}^F{}_W(t)\right] - \sum_\alpha \frac{P_\alpha}{M_\alpha} \frac{\partial \hat{\rho}^F{}_W(t)}{\partial R_\alpha} + \frac{1}{2}\sum_\alpha \left\{\frac{\partial \hat{V}^F{}_W(\boldsymbol{R},\hat{r},\hat{p})}{\partial R_\alpha}, \frac{\partial \hat{\rho}^F{}_W(t)}{\partial P_\alpha}\right\}. \tag{42}$$

Here, $\{,\}$ denotes the anti-commutator. We remind the authors that on the above Floquet QCLE, the density and Hamiltonian operators are now in the Floquet space.

The FQCLE will a platform for developing trajectory based nonadiabatic approach, such as Floquet surface hopping. Below, we show that we can further map the FQCLE into a Fokker-Planck equation when considering a manifold of electronic states. The resulting electronic friction represents the first order correction to the BO approximation.

### 3.2 Floquet electronic friction

Now, we show that we can map the FQCLE into a Fokker-Planck equation when the timescales of electronic and the driving are much faster than the nuclear motion. The deviation follows closely to the case of non Floquet driving [36, 37]. The nuclear density is given by $\mathcal{A}(\boldsymbol{R},\boldsymbol{P},t)=\frac{1}{N}Tr_{e,F}\left(\hat{\rho}^F{}_W(\boldsymbol{R},\boldsymbol{P},t)\right)$. Here we trace over all electronic DoFs as well as all Floquet replicas. Here N is the number of Floquet replicas. $Tr_{e,F}$ represent the trace over the Fourier space and many-body electronic states. To the first order correction to the BO approximation, we find that the nuclear motion follows the FP equation as follows:

$$\frac{\partial \mathcal{A}(t)}{\partial t} = -\sum_\alpha \frac{P_\alpha}{M_\alpha} \frac{\partial \mathcal{A}(t)}{\partial R^\alpha} - \sum_\alpha F_\alpha \frac{\partial \mathcal{A}(t)}{\partial P_\alpha} + \sum_{\alpha,\beta} \gamma_{\alpha\beta} \frac{\partial}{\partial P_\alpha}\left(\frac{P_\beta}{M_\beta}\mathcal{A}(t)\right) + \sum_{\alpha,\beta} \bar{\mathcal{D}}^S_{\alpha\beta} \frac{\partial^2 \mathcal{A}(t)}{\partial P_\alpha \partial P_\beta}, \qquad (43)$$

Here $F_\alpha$ is the mean force, $\gamma_{\alpha\beta}$ is the frictional tensor that represents the first order correction to the BO approximation. $\bar{\mathcal{D}}^S_{\alpha\beta}$ is the correlation function of the random force. In particular, we can express the frictional force as

$$\gamma_{\alpha\beta} = -\int_0^\infty dt\, Tr_{e,F}\left(\frac{\partial H^F}{\partial R_\alpha} e^{-iH^F t} \frac{\partial \hat{\rho}^F_{ss}}{\partial R_\beta} e^{iH^F t}\right), \qquad (44)$$

Here $\hat{\rho}^F_{ss}$ is the steady state Floquet electronic density operator. Thus, we have arrived in a general form for electronic friction in which all contributors are in their Floquet representations.

### 3.3 Numerical Results: Electronic friction

When applying the results for electronic friction to the model Hamiltonian in Equation (22), we can express the Floquet electronic friction in terms of Floquet Green's functions

$$\gamma_{\alpha\beta} = -\frac{1}{N}\int_{-\infty}^\infty \frac{d\epsilon}{2\pi} Tr_{m,F}\left(\frac{\partial h^F}{\partial R_\alpha} \frac{\partial G^{FR}(\epsilon)}{\partial \epsilon} \frac{\partial h^F}{\partial R_\beta} G^{F<}(\epsilon)\right) + h.c. \qquad (45)$$

Here $Tr_{m,F}$ refers to the trace over one-body and Floquet DoFs, and $h^F$ is the matrix elements of the System's Hamiltonian in the Floquet space. And $G^{FR}(\epsilon)$ and $G^{F<}(\epsilon)$ are Floquet retarded Green's function and Floquet lesser Green's function respectively. For more details consult with the Ref. [35].

To evaluate electronic friction given in terms the Floquet Green's functions, we choose a two-level electronic system carrying by two nuclear DoFs. Our model Hamiltonian for such system is given by:

$$h(\mathbf{R}, t) = \begin{pmatrix} x + \Delta & Ay + B\cos(\omega t) \\ Ay + B\cos(\omega t) & -x - \Delta \end{pmatrix} \qquad (46)$$

By taking the nuclear potential, $U(\mathbf{R})$, to be harmonic oscillators in both $x$ and $y$ dimensions, the diagonal elements of the total Hamiltonian represent two parabolas shifted in the x direction with a driving force of $2\Delta$. The off-diagonal coupling has two terms. The first one depends to the nuclear displacement in y direction with the strength $A$ and the second is an external time-periodic driving which represents the dipole approximation for a monochromatic light-matter interaction. $B$ and $\omega$ represent the intensity and frequency of the light. We have assumed one of levels connected to the left and another connected to the right leads such that $\Sigma^R_k$ is a diagonal matrix with the same elements as: $\Gamma_{11} = \Gamma_{22} = 1$. In addition, we have employed following variables: $A = 1, B = 2, \Delta = 3, \beta = 2$, and $\mu_L = \mu_R = $

0. Consistent with previous works [38], in the non-Floquet limit ($B = 0$), the antisymmetric part of electronic friction, $\gamma_{xy}^A = (\gamma_{xy} - \gamma_{yx})/2$, is vanishing, whereas the symmetric part, $\gamma_{xy}^S = (\gamma_{xy} + \gamma_{yx})/2$, is non-vanishing. In Figure 1 (a) and (b), we have plotted different electronic friction tensors, calculated by Equation (46), for $\omega = 1$ and $\omega = 0.5$, receptively. As one can see from these figures that, for $B \neq 0$, the $\gamma_{xy}^A$ is non-vanishing and its strength increases as $B$ increases.

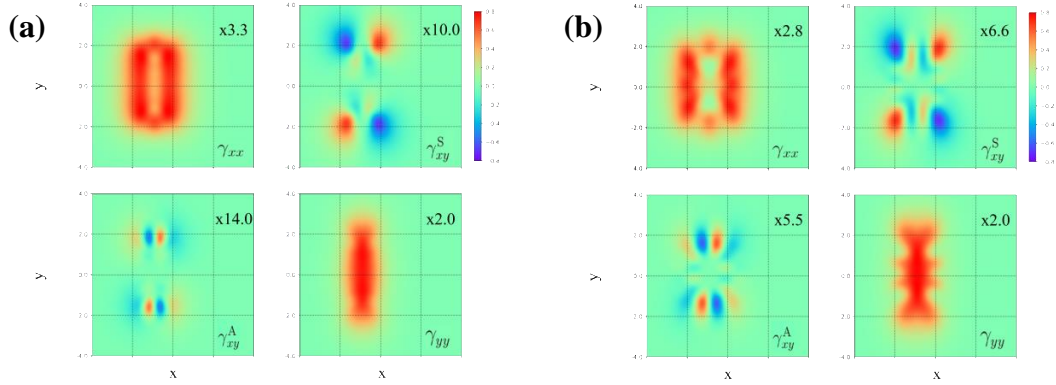

Figure 1. Electronic friction tensors for Floquet systems. (a) for $\omega = 1$ (b) for $\omega = 0.5$. Parameters are set to $\Gamma_{11} = \Gamma_{22} = 1$, $A = B = 1$, $\Delta = 2$, $\beta = 2$, and $\mu_L = \mu_R = 0$.

The nonvanishing $\gamma_{xy}^A$ results in a Lorentz-like force. Previously, we have shown that when including spin-orbit couplings, the friction tensor exhibits nonvanishing anti-symmetric parts.[38] Here, we show that the light-matter interactions can also introduce nonvanishing $\gamma_{xy}^A$. The nonvanishing Lorentz-like force is missing in the BO approximation, which may have a strong effect on nuclear dynamics [39]. Particularly, we have shown that such a Lorentz-like force can results in spin selectivity [39]

## 4. FLOQUET NONADIABATIC DYNAMICS IN OPEN QUANTUM SYSTEM

### 4.1 QAUNTUM-CLASSICAL LIOUVILLE EQUATION - CLASSICAL MASTER EQUATION (QCLE-CME)

When consider the nonadiabatic dynamics in open quantum system with Floquet driving, we can embed the FQCLE into a master equation. The procedure follows from the Floquet QME (Equation (36)) for the system, and then take the partial Wigner transformation on the nuclear DoFs. Finally, we arrive at FQCLE-CME:

$$\frac{\partial}{\partial t}\hat{\rho}_{S\,W}^F(\boldsymbol{R},\boldsymbol{P},t) = \{\hat{H}_{S\,W}^F(\boldsymbol{R},\boldsymbol{P}),\hat{\rho}_{S\,W}^F(\boldsymbol{R},\boldsymbol{P},t)\}_a - \mathrm{i}[\hat{H}_{S\,W}^F(\boldsymbol{R},\boldsymbol{P}),\hat{\rho}_{S\,W}^F(\boldsymbol{R},\boldsymbol{P},t)] - \widehat{\mathcal{D}}_W^F(\hat{\rho}_{S\,W}^F(\boldsymbol{R},\boldsymbol{P},t)) \quad (47)$$

Here, different from the FQCLE, we have an additional dissipator, $\widehat{\mathcal{D}}_W^F(\hat{\rho}_{S\,W}^F(\boldsymbol{R},\boldsymbol{P},t))$, given by

$$\widehat{\mathcal{D}}_W^F(\hat{\rho}_{S\,W}^F(\boldsymbol{R},\boldsymbol{P},t)) = -\int_0^\infty d\tau Tr_B\left([\widehat{\mathbb{H}}_{SB\,W}^F,[\widetilde{\widehat{\mathbb{H}}}_{SB\,W}^F(\tau),\hat{\rho}_{S\,W}^F(\boldsymbol{R},\boldsymbol{P},t)\otimes\hat{\rho}_B]]\right) \tag{48}$$

Notice that this dissipator is being partial Wigner transformed on the nuclear DoFs. We can further simplify this dissipator when considering the model Hamiltonian presented in Equation (22). The results will be similar to Equation (37) except that now the operator will be position dependent. The FQCLE-CME is a general approach to deal with Floquet nonadiabatic dynamics in open quantum systems. Below, we will apply this approach to a simple model.

**4.2 Anderson-Holstein model with Floquet driving: a case study**

We will now try to elaborate our methods based on the Anderson-Holstein model with Floquet driving. In our Anderson-Holstein model, there is only one level in the System, which couples to an electronic bath as well as a nuclear DoF. In addition, we apply a periodic driving to the electronic level. Such that the System Hamiltonian reads

$$\hat{H}_S(t) = \left(E_d + A\sin(\Omega t)\right)\hat{c}^\dagger\hat{c} + g(\hat{a}^\dagger + \hat{a})\hat{c}^\dagger\hat{c} + \hbar\omega\left(\hat{a}^\dagger\hat{a} + \frac{1}{2}\right) = \hat{H}_{mol} + A\sin(\Omega t)\hat{c}^\dagger\hat{c}. \tag{50}$$

In the System Hamiltonian $\hat{H}_S(t)$, which is separated into the time-independent part ($\hat{H}_{mol}$) and time-dependent part ($A\sin(\Omega t)\hat{c}^\dagger\hat{c}$), we consider the one level with the energy of $E_d$ is coupled to an oscillator with frequency $\omega$, the coupling strength can be tuned by $g$. Additionally, the amplitude $A$ determines the strength of the time periodic driving and $\Omega$ is the driving frequency. In this special case, we can derive a classical master equation to describe the coupled electron-nuclear dynamics for the system [40].

$$\frac{\partial \hat{P}_0(x,p)}{\partial t} = \frac{\partial \hat{H}_0(x,p)}{\partial x}\frac{\partial \hat{P}_0(x,p)}{\partial p} - \frac{\partial \hat{H}_0(x,p)}{\partial p}\frac{\partial \hat{P}_0(x,p)}{\partial x} - \gamma_{0\to 1}\hat{P}_0(x,p) + \gamma_{1\to 0}\hat{P}_1(x,p)$$

$$\frac{\partial \hat{P}_1(x,p)}{\partial t} = \frac{\partial \hat{H}_1(x,p)}{\partial x}\frac{\partial \hat{P}_1(x,p)}{\partial p} - \frac{\partial \hat{H}_1(x,p)}{\partial p}\frac{\partial \hat{P}_1(x,p)}{\partial x} + \gamma_{0\to 1}\hat{P}_0(x,p) - \gamma_{1\to 0}\hat{P}_1(x,p)$$

$$\tag{51}$$

Here $\hat{P}_0(x,p)$ and $\hat{P}_1(x,p)$ are the phase space density with the level being empty or occupied. $\gamma_{0\to 1}$ and $\gamma_{1\to 0}$ are the hopping rates that is given by:

$$\gamma_{0\to 1} = \Gamma\tilde{f}(\Delta V)$$

$$\gamma_{1\to 0} = \Gamma\left(1 - \tilde{f}(\Delta V)\right) \tag{52}$$

$$\Delta V = \hat{H}_1 - \hat{H}_0 = \sqrt{2}gx + E_d$$

Here, $\tilde{f}(\Delta V)$ refers to the Fermi function that is modified with Floquet replicas (Bessel-modified fermi function). In general, $\tilde{f}(\Delta V)$ is time dependent [40]. However, in the upcoming result and taking the time-scale separation, we employed the following time-averaged Bessel-modified fermi function:

$$\tilde{f}(\epsilon) = \sum_n \left[J_n\left(\frac{A}{\Omega}\right)\right]^2 f(\epsilon - n\Omega) \quad (53)$$

Where $f$ is the Fermi function. In fact, the rate equations given in Equation (51) is our CME for one-level where the Floquet theory is applied only within its dissipator term. We will employ the following surface hopping (SH) algorithm, to solve the above F-CME approximately. The algorithm is as follows,

1. Prepare the initial $x$ and $p$. Choose the active potential surface (say the state 0 surface).
2. Evolve $x$ and $p$ according to classical EOM for a time interval $\Delta t$ on the active surface.
3. Calculating the hopping rate $\gamma_{0\to 1}$ based on $\tilde{f}(\Delta V)$ and generate a random number $\zeta \in [0,1]$. If $\zeta < \gamma_{0\to 1}\Delta t$, the nuclei switch to the other surface without momentum rescaling. Otherwise, the nuclei stay on the same surface.
4. Repeat step 2 and 3 for the desired number of time steps.

Our Floquet surface hopping approach thus refers to solving Bessel-CME using SH (F-CME-SH). In the Figure 2, we present the transient electronic and nuclear dynamics of the one-level case. In particular, we plotted the nuclear kinetic energy denoted by $E_k$ and the expectation value of the number operator denoted by $N$ while the electrochemical potential kept at zero. Two methods are compared against each other, namely above F-CME-SH and the QCLE that is associated with Hilbert space QME (H-QME). Results are given for three different values for the driving amplitudes. Note that when the driving amplitude is small ($A = 0.001$), the nuclear kinetics can reach to the equilibrium temperature of $E_k = 1/2kT$, however, when increasing the driving amplitude ($A = 0.01, 0.02$), the effective temperature is different from the equilibrium temperature. This indicates that system cannot reach to thermodynamic equilibrium under high intensity driving. We have realized that employing the time-averaged Bessel-modified fermi function in Figure 2 (F-CME-SH) allows the dynamics to reach to the correct steady state in longer time while the dynamics will miss the small oscillation features that is associated with the external driving (not visible in the Figure 2).

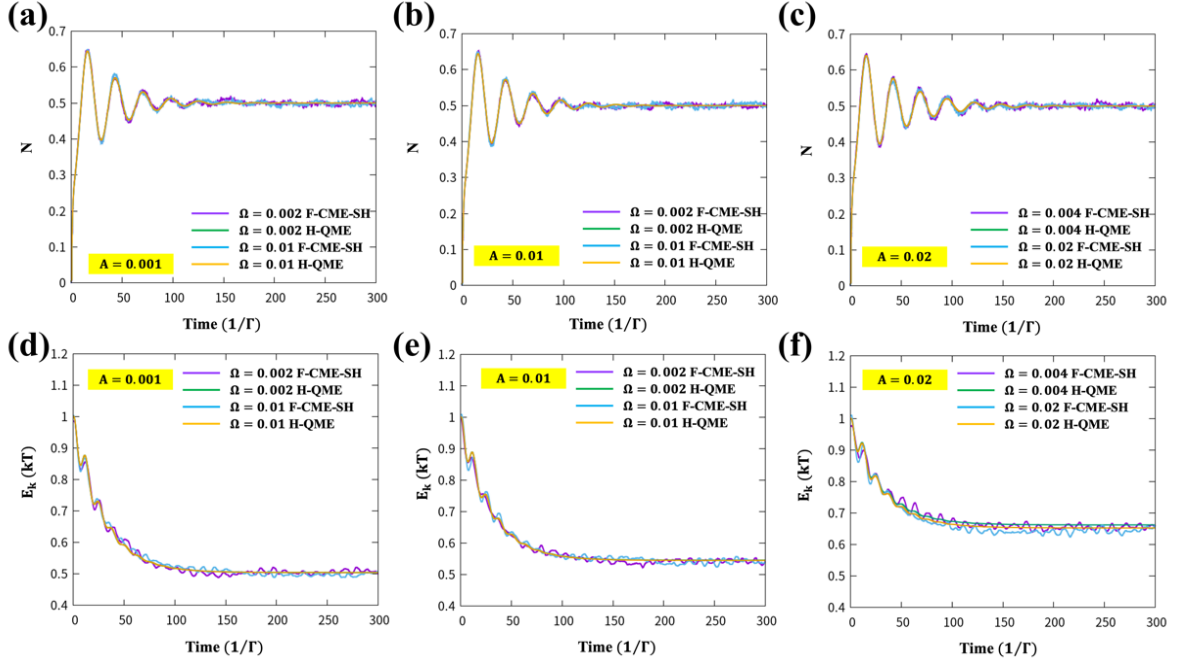

Figure 2. Transient dynamics: (a)-(c) the impurity electronic population as a function of time with different driving frequencies $\Omega$ and driving amplitudes $A$. (d)-(e) the nuclear relaxation as a function of time associated with (a)-(c). Here, we set nuclear frequency as $\hbar\omega = 0.003$, electron-phonon coupling $g = 0.0075$, $kT = 0.01$, $\Gamma = 0.01$, and $E_d = g^2/\hbar\omega$. Note that the effect temperature is different from equilibrium temperature ($E_k = 1/2kT$) when the driving amplitude is large ($A = 0.01, 0.02$).

**Conclusion**

In this overview, we offer Floquet formalism to treat dynamics, with a focus on nonadiabatic dynamics in open quantum systems. We begin by introduce Floquet LvN formulation, then two versions of Floquet QME, and Floquet QCLE. We also offer electronic friction approach, which is the first correction to the BO approximation with all electronic dynamics and Floquet driving being incorporated into frictional force. Finally, we sketch out Floquet QCLE-CME to treat nonadiabatic dynamics in open quantum systems in general. As an example, we consider a driven Anderson-Holstein model. We show that a Floquet surface hopping approach is applicable to treat the electronic and nuclear dynamics with periodic driving.

**Funding Information**

This work is supported by National Science Foundation of China under grant number NSFC No. 22273075.

**References**

1. Fiorin G, Klein ML, Hénin J. Using collective variables to drive molecular dynamics simulations. Molecular Physics. 2013;111(22-23):3345-62.
2. Ibrahim H, Lefebvre C, Bandrauk AD, Staudte A, Légaré F. H2: the benchmark molecule for ultrafast science and technologies. Journal of Physics B: Atomic, Molecular and Optical Physics. 2018;51(4):042002.
3. Pizzolato N, Fiasconaro A, Adorno DP, Spagnolo B. Resonant activation in polymer translocation: new insights into the escape dynamics of molecules driven by an oscillating field. Physical Biology. 2010;7(3):034001.
4. Bekard I, Dunstan DE. Electric field induced changes in protein conformation. Soft Matter. 2014;10(3):431-7.
5. L'Huillier A, Balcou P. High-order harmonic generation in rare gases with a 1-ps 1053-nm laser. Physical Review Letters. 1993;70(6):774.
6. Comstock M, Senekerimyan V, Dantus M. Ultrafast laser induced molecular alignment and deformation: Experimental evidence from neutral molecules and from fragment ions. The Journal of Physical Chemistry A. 2003;107(40):8271-81.
7. Essén H. The physics of the born–oppenheimer approximation. International Journal of Quantum Chemistry. 1977;12(4):721-35.
8. Duncan WR, Prezhdo OV. Theoretical studies of photoinduced electron transfer in dye-sensitized TiO2. Annu Rev Phys Chem. 2007;58:143-84.
9. Mandal A, Taylor M, Weight B, Koessler E, Li X, Huo P. Theoretical advances in polariton chemistry and molecular cavity quantum electrodynamics. 2022.
10. Zhu X, Xu Y, Zhao C, Jia C, Guo X. Recent Advances in Photochemical Reactions on Single-Molecule Electrical Platforms. Macromolecular Rapid Communications. 2022;43(16):2200017.
11. Chen H-T, Zhou Z, Subotnik JE. On the proper derivation of the Floquet-based quantum classical Liouville equation and surface hopping describing a molecule or material subject to an external field. The Journal of Chemical Physics. 2020;153(4):044116.
12. Wang H, Thoss M. Multilayer formulation of the multiconfiguration time-dependent Hartree theory. The Journal of chemical physics. 2003;119(3):1289-99.
13. Dou W, Bätge J, Levy A, Thoss M. Universal approach to quantum thermodynamics of strongly coupled systems under nonequilibrium conditions and external driving. Physical Review B. 2020;101(18):184304.
14. Bätge J, Levy A, Dou W, Thoss M. Nonadiabatically driven open quantum systems under out-of-equilibrium conditions: Effect of electron-phonon interaction. Physical Review B. 2022;106(7):075419.
15. Abedi A, Agostini F, Suzuki Y, Gross E. Dynamical steps that bridge piecewise adiabatic shapes in the exact time-dependent potential energy surface. Physical review letters. 2013;110(26):263001.
16. Curchod BF, Martínez TJ. Ab initio nonadiabatic quantum molecular dynamics. Chemical reviews. 2018;118(7):3305-36.
17. Subotnik JE. Fewest-switches surface hopping and decoherence in multiple dimensions. The Journal of Physical Chemistry A. 2011;115(44):12083-96.
18. Zhou Z, Chen H-T, Nitzan A, Subotnik JE. Nonadiabatic dynamics in a laser field: Using Floquet fewest switches surface hopping to calculate electronic populations for slow nuclear velocities. Journal of Chemical Theory and Computation. 2020;16(2):821-34.
19. Kuperman M, Nagar L, Peskin U. Mechanical stabilization of nanoscale conductors by plasmon oscillations. Nano Letters. 2020;20(7):5531-7.
20. Honeychurch TD, Kosov DS. Full counting statistics for electron transport in periodically driven quantum dots. Physical Review B. 2020;102(19):195409.
21. Schirò M, Eich FG, Agostini F. Quantum–classical nonadiabatic dynamics of Floquet driven systems. The Journal of Chemical Physics. 2021;154(11):114101.
22. Floquet G, editor Sur les équations différentielles linéaires à coefficients périodiques. Annales scientifiques de l'École normale supérieure; 1883.
23. Shirley JH. Solution of the Schrödinger equation with a Hamiltonian periodic in time. Physical Review. 1965;138(4B):B979.
24. Sambe H. Steady states and quasienergies of a quantum-mechanical system in an oscillating field. Physical Review A. 1973;7(6):2203.

25. Traversa FL, Di Ventra M, Bonani F. Generalized Floquet theory: Application to dynamical systems with memory and Bloch's theorem for nonlocal potentials. Physical Review Letters. 2013;110(17):170602.
26. Mahmood F, Chan C-K, Alpichshev Z, Gardner D, Lee Y, Lee PA, et al. Selective scattering between Floquet–Bloch and Volkov states in a topological insulator. Nature Physics. 2016;12(4):306-10.
27. Kundu A, Fertig H, Seradjeh B. Floquet-engineered valleytronics in dirac systems. Physical Review Letters. 2016;116(1):016802.
28. Kundu A, Fertig H, Seradjeh B. Effective theory of Floquet topological transitions. Physical review letters. 2014;113(23):236803.
29. Dou W, Subotnik JE. A generalized surface hopping algorithm to model nonadiabatic dynamics near metal surfaces: The case of multiple electronic orbitals. Journal of chemical theory and computation. 2017;13(6):2430-9.
30. Verzijl C, Seldenthuis J, Thijssen J. Applicability of the wide-band limit in DFT-based molecular transport calculations. The Journal of chemical physics. 2013;138(9):094102.
31. Wu B, Timm C. Noise spectra of ac-driven quantum dots: Floquet master-equation approach. Physical Review B. 2010;81(7):075309.
32. Kapral R. Progress in the theory of mixed quantum-classical dynamics. Annu Rev Phys Chem. 2006;57:129-57.
33. Kelly A, Kapral R. Quantum-classical description of environmental effects on electronic dynamics at conical intersections. The Journal of chemical physics. 2010;133(8):084502.
34. Subotnik JE, Ouyang W, Landry BR. Can we derive Tully's surface-hopping algorithm from the semiclassical quantum Liouville equation? Almost, but only with decoherence. The Journal of chemical physics. 2013;139(21):211101.
35. Mosallanejad V, Chen J, Dou W. Floquet driven frictional effects. arXiv preprint arXiv:220607894. 2022.
36. Dou W, Subotnik JE. A many-body states picture of electronic friction: The case of multiple orbitals and multiple electronic states. The Journal of Chemical Physics. 2016;145(5):054102.
37. Dou W, Miao G, Subotnik JE. Born-oppenheimer dynamics, electronic friction, and the inclusion of electron-electron interactions. Physical Review Letters. 2017;119(4):046001.
38. Teh H-H, Dou W, Subotnik JE. Spin polarization through a molecular junction based on nuclear Berry curvature effects. Physical Review B. 2022;106(18):184302.
39. Bode N, Viola Kusminskiy S, Egger R, von Oppen F. Scattering Theory of Current-Induced Forces in Mesoscopic Systems, Beilstein J. Nanotechnol. 2012;3:144.
40. Wang Y, Dou W. Nonadiabatic Dynamics Near Metal Surface With Periodic Drivings: A Floquet Surface Hopping Algorithm. arXiv preprint arXiv:230300479. 2023.